\documentclass[aps,twocolumn,pra,showpacs,amsmath,amssymb,showkeys,10pt]{revtex4-1}
\usepackage{graphicx}
\usepackage{epstopdf}
\usepackage{braket}
\usepackage{hyperref}
\allowdisplaybreaks

\begin{document}

\title{Cavity-field distribution in multiphoton Jaynes-Cummings resonances}

\author{Th. K. Mavrogordatos}
\email[Email address: ]{th.mavrogordatos@gmail.com}
\affiliation{Department of Physics, Stockholm University, SE-106 91, Stockholm, Sweden}

\date{\today}

\begin{abstract}
 We calculate the cavity-field distribution in the Wigner representation for the two-photon resonance of the weakly driven Jaynes-Cummings (JC) oscillator in its strong-coupling limit. Using an effective four-level system, we analytically demonstrate the presence of steady-state and transient bimodality which breaks azimuthal symmetry in phase space. The two steady-state peaks are located at opposite positions and do not correspond to the two-photon amplitude of the driven transition. The developing bimodality is portrayed in parallel with the evolution of the intensity correlation function for the forwards-scattered photons, before being finally contrasted to the few-photon steady-state and transient phase-space profiles for the cavity mode in the JC model driven on resonance.   
\end{abstract}


\maketitle

\section{Introduction}

The two-level behavior of an optical cavity mode strongly coupled to a single atom and excited near a vacuum Rabi resonance has been demonstrated about three decades ago as a paradigmatic system relying on the $\sqrt{n}$ nonlinearity of the Jaynes-Cummings (JC) model and exhibiting photon blockade~\cite{Tian1992}. This effect was reported a few years after single-atom absorptive optical bistability had been ascertained in a regime lying at the ``interface between the quantum limit, in which quantum-mechanical noise invalidates the semiclassical prediction of bistability, and the classical limit, in which quantum noise is a negligible perturbation on semiclassical results''~\cite{SingleAtomBist}. Only a handful of years until nowadays have lapsed, however, since the experimental observation of two-photon blockade in the strong-coupling limit of cavity QED~\cite{Hamsen2017} and the demonstration of the photon blockade breakdown in circuit QED by means of a first-order dissipative quantum phase transition~\cite{Fink2017}. This experimental verification of criticality in zero dimensions has been very recently followed by a theory for the critical scaling of steady-state observables as the critical point of the second-order phase transition associated with photon blockade~\cite{PhotonBlockade2015} is approached~\cite{Curtis2021}.  

Apart from its rather obvious link to photon antibunching and sub-Poissonian statistics~\cite{Carmichael1976, Kimble1977, Dagenais1978, Short1983, Birnbaum2005, Lang2011}, which arise as distinct quantum features, photon blockade has also been associated with a more intricate property, namely the existence of a strong-coupling ``thermodynamic limit'' in single-atom QED, where the relevant system size parameter -- revealed for a vanishing spontaneous emission rate -- is proportional to the square of the light-matter coupling strength. Upon approaching this limit, quantum fluctuations are not a``negligible perturbation on semiclassical results''. Instead, multiphoton resonances of higher order appear, turning increasingly sharper before they saturate for a growing drive amplitude, while the discrepancy from the semiclassical nonlinear response persists (see also Sec. IV-C of~\cite{PhotonBlockade2015}). 

In their experimental report on the properties of the two-photon resonance for one trapped atom in a high-finesse optical cavity, Schuster and collaborators mention that they ``also analyzed the nonlinear theory of optical bistability and found that it is inconsistent with all of the measurements presented'', identifying the operation on the lower branch of bistability where the nonlinear response is weak~\cite{Schuster2008}. Shortly after, in one of the exemplary experiments of circuit QED, spectroscopic evidence for multiphoton dressed states along the JC ladder was provided by heterodyne transmission measurements in a setup using a superconducting qubit coupled to a microwave field~\cite{Bishop2009}. Therein, Bishop and collaborators appeal to an effective two-level system to model the supersplitting of the vacuum Rabi resonance, bringing us back to the early ideas on photon blockade developed in~\cite{Tian1992} and their two-mode extension implemented in the experiment of~\cite{Birnbaum2005}.

Bringing these pieces together, in our work we aim to bridge the gap between the aforementioned quantum and classical limits focusing on a multiphoton resonance, a configuration where the classical nonlinearity is intertwined with quantum fluctuations in a fundamental way that is pictured in phase space. For the special case of the two-photon transition the link is provided by low-amplitude quantum bistability, or rather bimodality, evidenced in the {\it quasi}probability distribution of the intracavity field in the strong-coupling regime. After reducing our description to a minimal four-level model in Sec.~\ref{sec:effMEf}, we first calculate the steady-state Wigner function at {\it quasi}resonant excitation of the transition, and we then relate the transient bimodality to the evolution of the intensity correlation function for the forwards scattered photons in Sec.~\ref{sec:distcavcoh}. We then find in Sec.~\ref{sec:secapproxrev} that the multi-modality accompanying a multiphoton resonance disappears when the JC oscillator is resonantly excited with a weak drive amplitude. Instead, in the nonperturbative treatment one speaks of two counter-propagating wavepackets in close connection with the initial Gaussian shape of the state naturally produced by the coherent driving. In the last part of our discussion, reserved for the Conclusions, we point to a possible experimental observation of the change occurring in the multiphoton quantum nonlinearity when we transition from the photon-blockade regime to resonant excitation.  

\section{Formulating the effective master equation}
\label{sec:effMEf}

The system density matrix $\rho$ for the open driven JC model obeys the Lindblad master equation (ME)
\begin{align}\label{eq:ME1}
 \frac{d\rho}{dt}&=-i[\omega_0(\sigma_{+}\sigma_{-} + a^{\dagger}a)+g(a\sigma_{+}+a^{\dagger}\sigma_{-}),\rho]\notag \\
 &-i[ (\varepsilon_d^{*} a e^{i\omega_d t} + \varepsilon_d a^{\dagger}e^{-i\omega_d t}),\rho]\notag \\
 &+\kappa (2 a \rho a^{\dagger} -a^{\dagger}a \rho - \rho a^{\dagger}a)\notag \\
 &+\frac{\gamma}{2}(2\sigma_{-}\rho \sigma_{+} - \sigma_{+}\sigma_{-}\rho - \rho \sigma_{+}\sigma_{-}),
\end{align}
where $a$ and $a^{\dagger}$ are the annihilation and creation operators for the cavity photons, $\sigma_{+}$ and $\sigma_{-}$ are the raising and lowering operators for the two-level atom, $g$ is the dipole coupling strength, $2\kappa$ is the photon loss rate from the cavity, and $\gamma$ is the spontaneous emission rate for the atom to modes other than the privileged cavity mode which is coherently driven with amplitude $\varepsilon_d$ and frequency $\omega_d$ defining the detuning $\Delta \omega_d \equiv \omega_d-\omega_0$. Hereinafter, we consider the special case $\gamma=2\kappa$ which considerably simplifies the analysis, while we take $g/\kappa=1000$ after~\cite{Shamailov2010, Bishop2009}. 

We now employ an effective four-level model to produce an analytical expression for the matrix elements of the system density matrix when the two-photon transition is resonantly excited in the weak excitation regime. Such a perturbative treatment, devised in~\cite{Shamailov2010} to study the first and second-order coherence of the forwards scattered light for the two-photon resonance in one-atom cavity QED, uses essentially an expansion in powers of $(\varepsilon_d/g)^2$ (without loss of generality for this derivation we assume that the driving field amplitude $\varepsilon_d$ is real). The first four JC dressed states between which transitions are constrained to occur, starting from the ground state, are denoted by
\begin{subequations}\label{eq:4levels}
\begin{align}
 &\ket{\xi_0}\equiv\ket{0,-}, \\
 &\ket{\xi_1}\equiv \frac{1}{\sqrt{2}} (\ket{1,-} - \ket{0,+}), \\
 & \ket{\xi_2}\equiv \frac{1}{\sqrt{2}} (\ket{1,-} + \ket{0,+}), \\
 &  \ket{\xi_3}\equiv \frac{1}{\sqrt{2}} (\ket{2,-} - \ket{1,+}),
\end{align}
\end{subequations}
where $\ket{n, \pm}\equiv \ket{n}\otimes\ket{\pm}$, $\ket{n}$ is the Fock state of the cavity field, while $\ket{+}, \ket{-}$ are the upper and lower states of the two-level atom, respectively.
\begin{figure}
\centering
\includegraphics[width=0.4\textwidth]{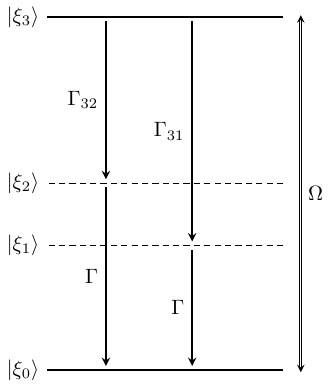}
\caption{{\it The minimal four-level model.} Schematic diagram of the four dressed-state levels with energies given by Eqs.~\eqref{eq:dressedenergies} (see Fig. 2 of~\cite{Shamailov2010}). The two-photon Rabi frequency $\Omega$ scales with the square of the drive amplitude to dominant order in the perturbative expansion, and the transition rates are given by Eqs.~\eqref{eq:rates}. The dynamical evolution for this description is governed by the effective ME~\eqref{eq:ME2}. The intermediate levels $\ket{\xi_1}, \ket{\xi_2}$, marked by dashed lines in the diagram, are occupied via the cascaded decay.}
\label{fig:leveldiagram}
\end{figure}
The secular approximation in the limit of strong nonperturbative coupling ($g \gg \kappa, \gamma/2$) leads to the following effective ME in the laboratory frame [see~\cite{Lledo2021} for further details, and Eq. (18) of~\cite{Shamailov2010}]
\begin{equation}\label{eq:ME2}
\begin{aligned}
  &\frac{d\rho}{dt}=\mathcal{L}\rho \equiv -i[\tilde{H}_{\rm eff},\rho]+\Gamma_{32} \mathcal{D}[|\xi_2\rangle \langle \xi_3|](\rho)\\
  & + \Gamma_{31} \mathcal{D}[|\xi_1\rangle \langle \xi_3|](\rho)+ \Gamma \mathcal{D}[|\xi_0\rangle \langle \xi_1|](\rho) + \Gamma \mathcal{D}[|\xi_0\rangle \langle \xi_2|](\rho)
  \end{aligned}
\end{equation}
where the effective Hamiltonian of the four-state model is
\begin{equation}\label{eq:EffHam}
 \tilde{H}_{\rm eff}\equiv\sum_{k=0}^{3} \tilde{E}_{k} |\xi_k\rangle \langle \xi_k| + \hbar \Omega (e^{2i\omega_d t} |\xi_0\rangle \langle \xi_3| + e^{-2i\omega_d t} |\xi_3\rangle \langle \xi_0|),
\end{equation}
with the following energies (shifted by $\delta_0$--$\delta_3$) for the four states dressed by the drive,
\begin{subequations}\label{eq:dressedenergies}
\begin{align}
 &\tilde{E}_0=E_0 + \hbar\delta_0(\varepsilon_d) = \hbar \sqrt{2} \varepsilon_d^2/g, \label{eq:E0shift}\\
 & \tilde{E}_1=E_1 + \hbar\delta_1(\varepsilon_d) = \hbar \{\omega_0 -  g - [(20 + 19\sqrt{2})/7]\varepsilon_d^2/g\}, \label{eq:E1shift} \\
 & \tilde{E}_2=E_2 + \hbar\delta_2(\varepsilon_d) = \hbar \{\omega_0 +  g +  [(20 - 19\sqrt{2})/7]\varepsilon_d^2/g\}, \label{eq:E2shift} \\
 & \tilde{E}_3=E_3 + \hbar\delta_3(\varepsilon_d) =  \hbar (2\omega_0 - \sqrt{2}  g -  \sqrt{2}\, \varepsilon_d^2/g), \label{eq:E3shift} 
\end{align}
\end{subequations}
and $\Omega =2 \sqrt{2} \varepsilon_d^2/g$ the two-photon Rabi frequency to dominant order~\cite{Lledo2021}. In the effective ME of Eq.~\eqref{eq:ME2}, we define as usual $\mathcal{D}[X](\rho)\equiv X\rho X^{\dagger}-(1/2)\{X^{\dagger}X, \rho\}$, while
\begin{subequations}\label{eq:rates}
\begin{align}
 &\Gamma_{31}\equiv\frac{\gamma}{4}+(\sqrt{2}+1)^2 \frac{\kappa}{2}=\frac{\gamma}{4}[1+(\sqrt{2}+1)^2], \\
 & \Gamma_{32}\equiv\frac{\gamma}{4}+(\sqrt{2}-1)^2 \frac{\kappa}{2}=\frac{\gamma}{4}[1+(\sqrt{2}-1)^2], \\
 & \Gamma\equiv\frac{\gamma}{2}+\kappa=\gamma,
\end{align}
\end{subequations}
are the transition rates between the dressed states. The shifted dressed-state levels and the transition rates between them comprise the effective four-level model depicted in Fig.~\ref{fig:leveldiagram}. From Eqs.~\eqref{eq:E0shift} and~\eqref{eq:E3shift} we find that the two-photon resonance, including the level shifts, must be excited with a drive frequency $\omega_d$ given by
\begin{equation}
 2\omega_d=(\tilde{E}_3-\tilde{E}_1)/\hbar=2\omega_0-\sqrt{2}g + \delta_3-\delta_0,
\end{equation}
including second-order corrections with respect to the drive amplitude.

\section{From the cavity-field distribution to the coherence of the forwards-scattered photons}
\label{sec:distcavcoh}

The effective ME~\eqref{eq:ME2} can be now used to evolve the system density matrix from a given initial state. For a general time $\tau$, the reduced density matrix for the cavity field, defined as $\rho_c \equiv \braket{+|\rho|+} + \braket{-|\rho|-}$ , can be written in the form
\begin{equation}\label{eq:rhocevolv}
\begin{aligned}
  \rho_{c}(\tau)&=\left[\rho_{00}+\frac{1}{2}(\rho_{11}+\rho_{22})-{\rm Re}(\rho_{12}) \right] |0\rangle \langle 0| \\
  &+\left[\frac{1}{2}(\rho_{11}+\rho_{22}+\rho_{33})+{\rm Re}(\rho_{12}) \right] |1\rangle \langle 1|\\
  &+ \frac{1}{2}\rho_{33} |2\rangle \langle 2| +\frac{1}{\sqrt{2}}\left(\rho_{03} |0\rangle \langle 2| +  \rho_{03}^{*} |2\rangle \langle 0|\right),
\end{aligned}
\end{equation}
where the matrix elements are taken with respect to the dressed states $\ket{\xi_0}$--$\ket{\xi_3}$, while we omit their time arguments for brevity. The time dependence of the matrix elements featuring in the above expression can be calculated as follows, starting from
\begin{equation}
\begin{aligned}
&\rho_{33}=C e^{-2\gamma\tau} + \frac{\Omega^2}{\gamma^2 + 4\Omega^2}[1-(\gamma/\Omega)e^{-\gamma \tau}\sin(2\Omega \tau)] \\
&- \Sigma e^{-\gamma \tau} \frac{\Omega}{\gamma^2 + 4\Omega^2}[\gamma \sin(2\Omega \tau)-2\Omega \cos(2\Omega \tau)],
 \end{aligned}
\end{equation}
featuring two constants to be determined from the initial conditions with $\Sigma \equiv\rho_{33}(0)-\rho_{00}(0)$ and $C$ expressed in terms of $\Sigma$ via $\rho_{33}(0)$. We then go on to determine the sum of occupation probabilities
\begin{equation}
 \rho_{11} + \rho_{22}= C^{\prime} e^{-\gamma\tau} + 2\gamma e^{-\gamma\tau} \int \rho_{33}\, e^{\gamma\tau}d\tau 
\end{equation}
comprising yet another constant to be determined from $\rho_{11}(0) + \rho_{22}(0)$. Now, $\rho_{00}$ follows from normalization, while the off-diagonal elements involved read
\begin{subequations}
\begin{align}
 & \rho_{12}=\rho_{21}^{*}=\rho_{12}(0) e^{-\Gamma \tau} e^{i \nu \tau}, \label{eq:qbeat}\\
 & \rho_{03}=\rho_{30}^{*}=\frac{i\Omega \gamma}{\gamma^2 + 4\Omega^2} -i \frac{1}{2}\Sigma e^{-\gamma \tau} \sin(2\Omega \tau)\notag \\
  &-i\frac{1}{2}\frac{\gamma}{\gamma^2+4\Omega^2} e^{-\gamma \tau} [\gamma\sin(2\Omega \tau)+2\Omega\cos(2\Omega \tau)].
\end{align}
\end{subequations}
The two complex conjugate matrix elements in Eq.~\eqref{eq:qbeat} generate a quantum beat at the frequency $\nu=2g + \delta_2-\delta_1$ [see Eqs.~\eqref{eq:E1shift} and~\eqref{eq:E2shift}, as well as Eq.~\eqref{eq:g2cav}] which is significantly higher than all the other rates involved in the dynamics, and is revealed by the intensity correlation function of the transmitted light, as we will see later on. With these results in hand, we can now propagate to the steady state. Our final expressions for the steady state are given in terms of the excitation probability of level $\ket{\xi_3}$ -- the upper level of the two-photon transition -- denoted by $p_3=\Omega^2/(\gamma^2 + 4\Omega^2)$~\cite{Shamailov2010}. 

The Wigner distribution $W(\alpha, \alpha^{*})$ (in the complex variable $\alpha=x + iy$ and its conjugate, $\alpha^{*}=x-iy$) is the Fourier transform of the symmetrically-ordered characteristic function (see Ch. 4 of~\cite{CarmichaelQO1})
\begin{equation}
\chi_S(z,z^{*}) \equiv {\rm tr} \left( \rho e^{iz^{*}a^{\dagger}+iza} \right), 
\end{equation}
and is defined as
\begin{equation}
\begin{aligned}
&W(\alpha, \alpha^{*}) \equiv \frac{1}{\pi^2} \int d^2 z\, \chi_S(z,z^{*})\, e^{-iz^{*}\alpha^{*}}\, e^{-iz\alpha}\\
&=  \frac{1}{\pi^2} \int_{-\infty}^{\infty}\, d\mu \int_{-\infty}^{\infty} d\nu \, \chi_S (\mu + i\nu, \mu - i\nu)\, e^{-2i(\mu x -\nu y)}.
\end{aligned}
\end{equation}
A central result to guide our discussion is the normalized Wigner distribution of the (intra)cavity field in the steady state, which we derive in the form 
\begin{equation}\label{eq:Wss}
\begin{aligned}
 W_{\rm ss}(\alpha, \alpha^{*})&=\frac{2}{\pi}e^{-2|\alpha|^2}\Big\{4p_3 |\alpha|^4 +2 p_3 |\alpha|^2 +(1 -3 p_3)\\
 &+ i 2\, \sqrt{p_3(1-4p_3)}\,[\alpha^2-(\alpha^{*})^2]\Big\}.
\end{aligned}
\end{equation}
\begin{figure*}
\centering
\includegraphics[width=\textwidth]{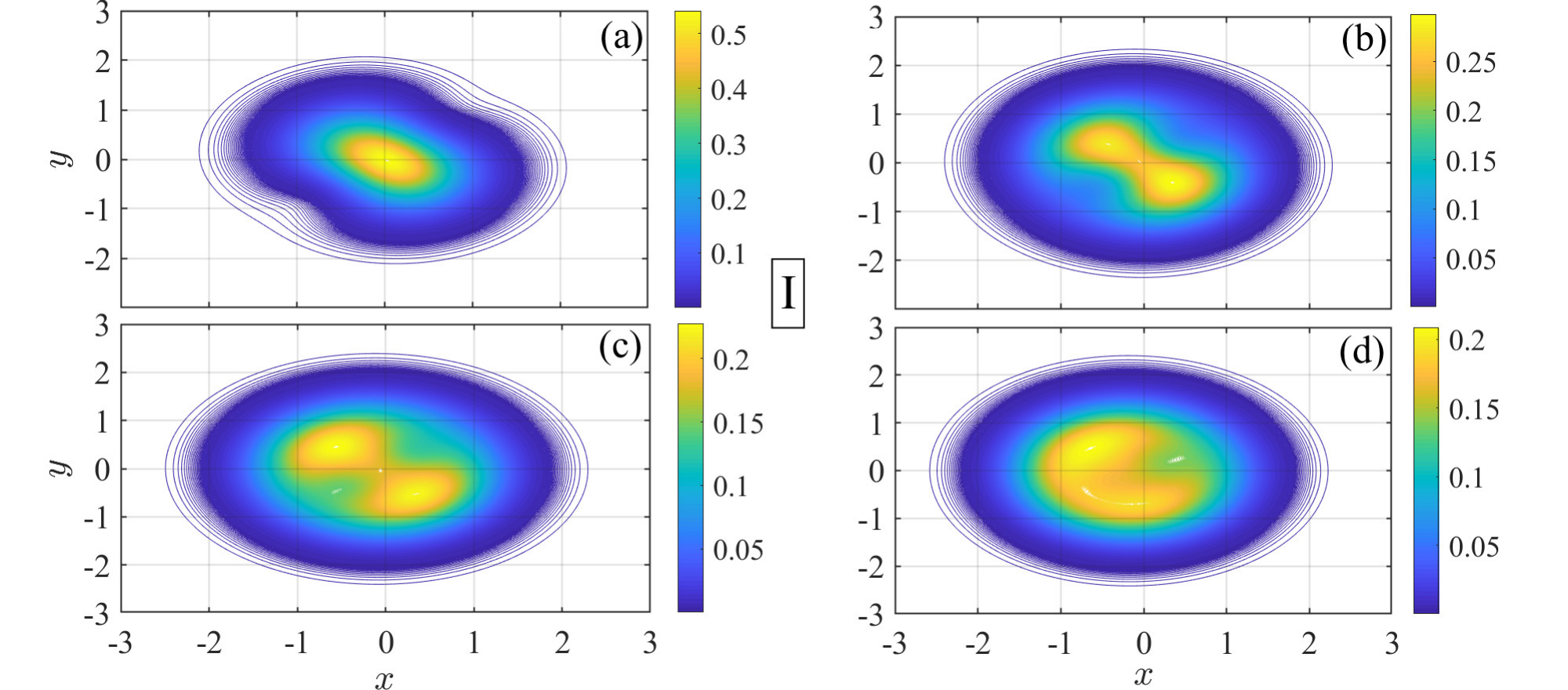}
\includegraphics[width=\textwidth]{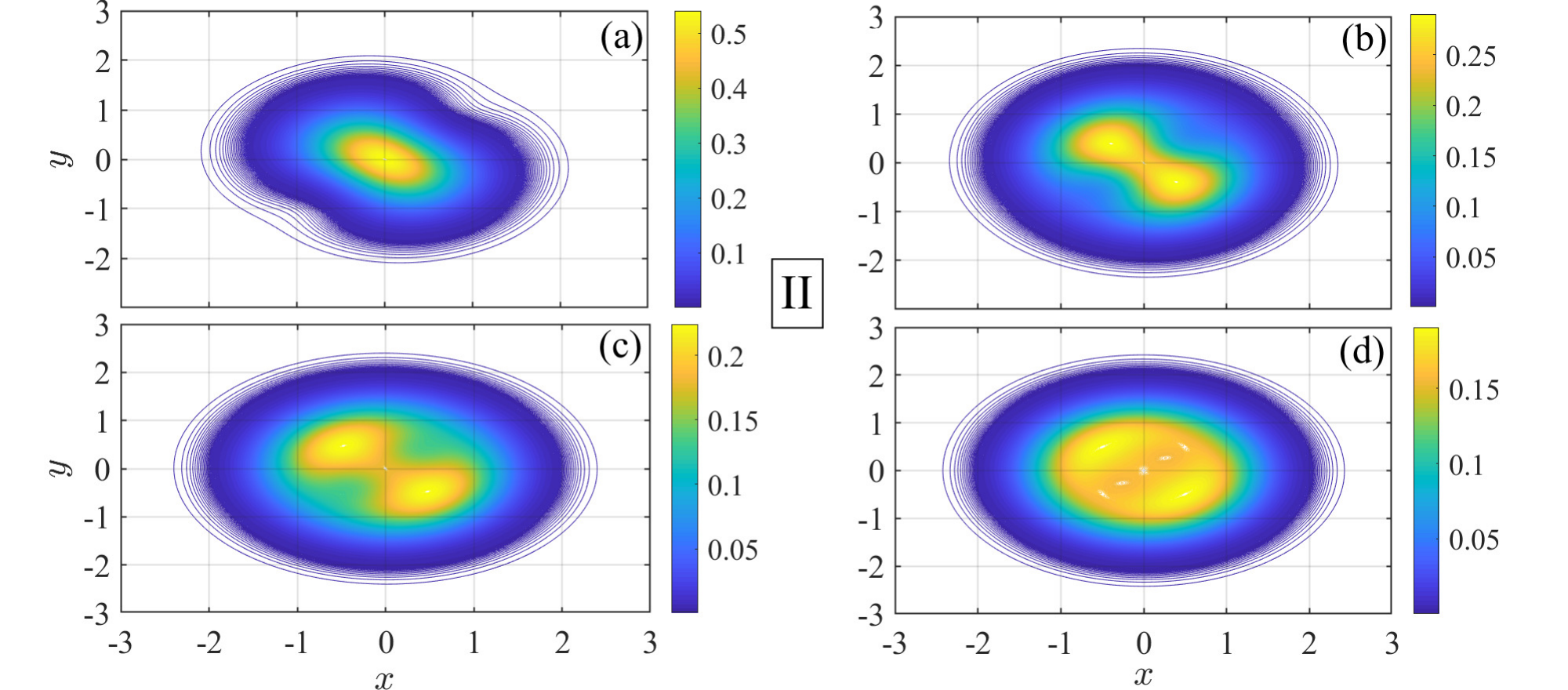}
\caption{{\it Steady-state cavity-field distributions.} Wigner function (contour plots) of the intracavity field, $W_{\rm ss}(x+iy)$, extracted from the steady-state solution of the ME~\eqref{eq:ME1}, depicted in {\bf Panel I}, against the Wigner functions computed from Eq.~\eqref{eq:Wss} and depicted in {\bf Panel II}, for $\Delta \omega_d=-g/\sqrt{2}-\sqrt{2}\varepsilon_d^2/g$ and $g/\gamma=500$. Frames {\bf (a)}--{\bf (d)} in each panel show distributions plotted for $p_3= 0.05, 0.2, 0.24, 0.249$, respectively.  The numerical solution of the ME diagonalizes the Liouvillian for a Hilbert space of $30$ Fock states in Matlab's {\it Quantum Optics Toolbox}.}
\label{fig:WssAN}
\end{figure*}
The two terms featuring in the last line on the RHS are responsible for breaking the symmetry arising from a linear combination of the vacuum-state and Fock-state ($\ket{l}, l=0, 1, 2$) phase-space distributions (see Ch. 3 of~\cite{CarmichaelQO1}). Symmetry breaking originates from the off-diagonal element $|0 \rangle \langle 2|$ and its Hermitian conjugate in $\rho_c$, coming directly from the driving terms in the Hamiltonian~\eqref{eq:EffHam}. As the two-photon Rabi frequency increases, the steady-state distribution of Eq.~\eqref{eq:Wss} develops two peaks along the line $y=-x$ and two dips along the line $y=x$ in phase space (clearly forming for $p_3 \gtrsim 0.15$). The steady-state Wigner function remains everywhere positive for $0\leq p_3 \leq 1/4$.  

In Fig.~\ref{fig:WssAN}, we trace the onset of bimodality for increasing $p_3$ with reference to our effective four-level model, depicted in Fig.~\ref{fig:leveldiagram}, to which the dynamical evolution is constrained. 
\begin{figure*}
\centering
\includegraphics[width=\textwidth]{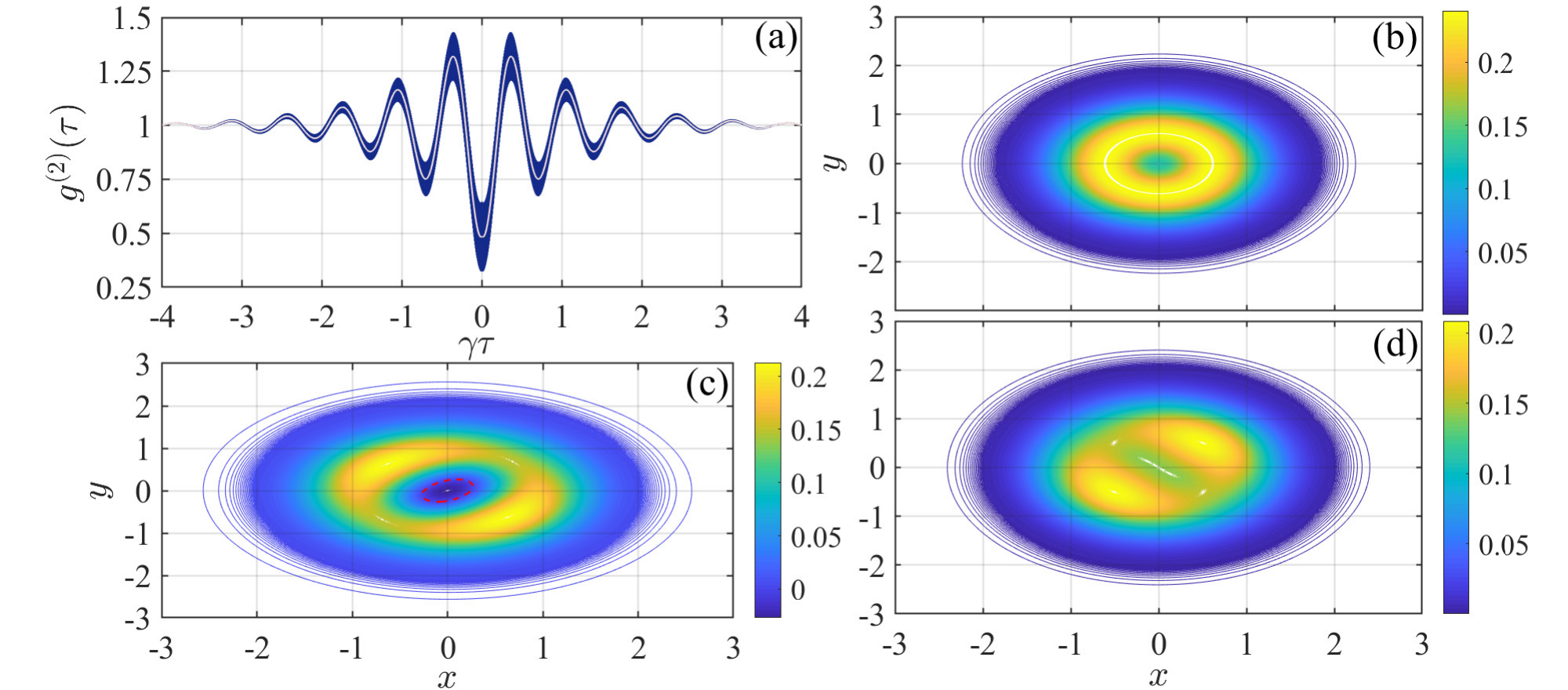}
\caption{{\it Second-order coherence and bimodality.} {\bf (a)} The intensity correlation function for the photons scattered in the forwards direction is plotted from Eq.~\eqref{eq:g2cav} with $p_3=0.247$. The solid white line averages over the quantum beat. {\bf (b)}--{\bf (d)} Contour plots of the Wigner functions $W_{\tau}(x+iy)$ which are analytically determined from the time-varying density matrix $\rho_c (\tau)$ of Eq.~\eqref{eq:rhocevolv} [extracted from the partial trace of $\rho(\tau)$ evolving from $\rho(0)=\rho_{\rm cond}(0)$] at the scaled times $\gamma \tau=0, 0.3549, 0.5401$, respectively. At the middle time in the sequence [corresponding to the distribution depicted in frame (c)] the correlation function attains its maximum, with $g^{(2)}(\gamma\tau=0.3549) \approx 1.43$, while $g^{(2)}(\gamma\tau=0.5401)=1$. The dashed line in frame (c) marks out the region inside which the Wigner function attains negative values.}
\label{fig:g2W}
\end{figure*}
The analytical expression for the Wigner function suggests that azimuthal symmetry is restored asymptotically as the two-photon transition saturates and the occupation probability $p_3$ tends to $1/4$ -- its maximal value. In fact, this picture reflects the limitations of the four-level model: as we can observe in Panel I of Fig.~\ref{fig:WssAN}, which depicts numerical results for the steady-state intracavity field distribution, the two peaks already deviate from the line $y=-x$ for $p_3=0.24$ (being located at $\alpha_{m_1} \approx 0.35 - 0.53 i$ and $\alpha_{m_2}\approx-0.56 +0.45 i$ in contrast to the analytical distribution placing the peaks at $\alpha_{m_{1,2}}\approx\pm 0.48 \mp 0.48i$). Agreement between exact numerical results and the analytical distributions shown in Panel II is also inherently compromised by the approximations involved in deriving Eq.~\eqref{eq:ME2} from~\eqref{eq:ME1}. Upon a further increase of the drive amplitude, when the perturbative expansion breaks down and higher levels along the JC ladders enter into play, the two peaks approach each other, as we can observe in frame (d) of Panel I, until they eventually merge into a single broad squeezed distribution. Qualitatively, we can expect that the $n$-photon resonance will bring about a multimodality of order $n$ through the term proportional to $[\alpha^n-(\alpha^{*})^n]e^{-2|\alpha|^2}$. For $\Delta \omega_d \approx -g/\sqrt{3}$, numerical simulations confirm the existence of three peaks in phase space with neighboring position vectors forming an angle of $2\pi/3$ between them. This is in agreement with the steady-state Wigner functions given in Fig. 8 of~\cite{Miranowicz2013} for the $n$-photon blockade encountered in the driven dissipative Kerr oscillator; the distribution exhibits $n$ peaks and $n$ dips. 

The Wigner function of Eq.~\eqref{eq:Wss} can be used for the calculation of various moments of the cavity field in the steady state, respecting symmetric operator ordering. For example, we compute the steady-state average photon number as
\begin{equation}
\begin{aligned}
&\left(\overline{\alpha^{*} \alpha}\right)_{W_{\rm ss}}=\int d^2\alpha\, W_{\rm ss} (\alpha, \alpha^{*})\alpha^{*}\alpha\\
&=\frac{1}{2}(1+5p_3)=\frac{1}{2}\left(2 \braket{a^{\dagger}a}_{\rm ss} +1 \right),
\end{aligned}
\end{equation}
whence we verify that $\braket{a^{\dagger}a}_{\rm ss}= 5 p_3/2$. Only the first three terms in Eq.~\eqref{eq:Wss}, those originating from the vacuum state and the Fock states with $l=1, 2$, contribute to the final result -- the symmetry-breaking terms multiplied by $|\alpha|^2$ drop out. 

Phase-space bimodality is also present in the transient regime. Following~\cite{Shamailov2010}, we select the initial density matrix as the conditional state occurring at the emission of one photon from the cavity after the attainment of steady state,
\begin{equation}\label{eq:condstate}
 \rho_{\rm cond}(0)=\frac{2}{5}|\xi_0 \rangle \langle \xi_0 | + \frac{3}{5} |\psi_b \rangle \langle \psi_b|,
\end{equation}
with the superposition generating a quantum beat of partial visibility,
\begin{equation}
 \ket{\psi_b}=\sqrt{\frac{2}{3}} \left(\frac{\sqrt{2}+1}{2}\ket{\xi_1} +  \frac{\sqrt{2}-1}{2}\ket{\xi_2} \right).
\end{equation}
This particular choice determines the set of constants
\begin{equation}
 \Sigma=-\frac{2}{5}, \quad C=-\frac{1}{5} p_3, \quad C^{\prime}=0, \quad \rho_{12}(0)=\frac{1}{10},
\end{equation}
and ultimately the time-varying density matrix from which we derive the {\it quasi}probability distribution of the intracavity field. It also determines the second-order correlation function for the forwards-scattered light, given in~\cite{Shamailov2010} as
\begin{equation}\label{eq:g2cav}
\begin{aligned}
 g^{(2)}(\tau)&=1+e^{-\gamma|\tau|}[a_1 \cos(2\Omega \tau) + a_2 \sin(2\Omega |\tau|)\\
 &+ a_3 e^{-\gamma|\tau|}+a_4 \cos(\nu \tau)],
 \end{aligned}
\end{equation}
with coefficients depending solely on $p_3$ through the ratio $\gamma/\Omega$:
\begin{subequations}
 \begin{align}
&a_1=3\frac{\gamma^2-4\Omega^2}{25\Omega^2}=\frac{3}{25p_3}(1-8p_3),\\
&a_2=-\frac{13}{25}\frac{\gamma}{\Omega}=-\frac{13}{25}\sqrt{\frac{1-4p_3}{p_3}},\\
&a_3=-\frac{1}{25},\\
&a_4=\frac{\gamma^2 + 4\Omega^2}{25\Omega^2}=\frac{1}{25p_3}.
 \end{align}
\end{subequations}
The quantum beat -- the term with coefficient $a_4$ in Eq.~\eqref{eq:g2cav} -- is averaged out in Fig.~\ref{fig:g2W}(a), leaving behind an oscillation whose period is determined by the two-photon Rabi frequency here scaled to $\Omega/\gamma \approx 4.54$ (corresponding to $\varepsilon_d/\gamma \approx 28.32$). In Figs.~\ref{fig:g2W}(b)--(d), we observe the evolution of the Wigner function from zero time delay to the time when $g^{(2)}(\tau)$ attains its maximum value. We encounter transient symmetry breaking, with the peaks of the cavity distribution placed at different positions along the lines $y=\pm x$ in phase space, always at equal distances from the origin. The conditional state of Eq.~\eqref{eq:condstate} has a circularly symmetric Wigner distribution, as we can see in Fig.~\ref{fig:g2W}(b), which is subsequently broken in the early stages of the evolution. It corresponds to photon antibunching, since $g^{(2)}(0)\approx 0.65<1$ and the subsequent evolution has a positive slope. The depicted curve satisfies the criterion ${g^{(2)}}^{\prime}(0)=0, \quad {g^{(2)}}^{\prime\prime}(0)>0$~\cite{Dagenais1978, Short1983}. Interestingly, the form of Eq.~\eqref{eq:g2cav} predicts photon bunching instead at low drive amplitudes (whence for low $p_3$), as noted in~\cite{Shamailov2010}. The maximum value of $g^{(2)}(\tau)$ is associated with a bimodal profile comprising two peaks separated by a region around the origin where the Wigner function is negative [Fig.~\ref{fig:g2W}(c)]---a remaining sign of the single-photon distribution originating from the dressed states $\ket{\xi_1}$ and $\ket{\xi_2}$ involved in the quantum beat. We note that the region separating the two peaks is directly affected by the period of the beat during the evolution ($2\pi/\nu$). For instance, in the Wigner function plotted for $\gamma\tau=0.3580$, half a beating period away from the location of the maximum of $g^{(2)}(\tau)$, the two peaks are joined by a pronounced ridge, and the distribution is everywhere positive. Finally, we find that the bimodality along the line $y=x$ [Fig.~\ref{fig:g2W}(d)] occurs during the time intervals when ${\rm Im}(\rho_{03})<0$. The existence of such intervals indicates a saturated two-photon transition ($p_3 > 0.24$). 

\section{JC interaction driven on resonance: from the secular approximation to quantum-state revivals}
\label{sec:secapproxrev}

Let us now move to the weak-excitation limit on resonance, where the evolution of the system density involves primarily the first three levels of the JC ladder $\ket{\xi_0}$, $\ket{\xi_1}$ and $\ket{\xi_2}$, according to the perturbative analysis of Sec. 16.3.4 in~\cite{CarmichaelQO2} within the secular approximation relying once more on what has been termed ``dressing of the dressed states''. In this weak-excitation regime, to dominant order, the Wigner function of the intracavity field has azimuthal symmetry with a leading contribution from the vacuum state and subsequently from the the Fock states $|l\rangle \langle l|$ with $l= 1, 2$. The corresponding coefficients of these three terms are expanded in powers of the squeezing parameter $r\equiv (|\varepsilon_d|/g)^2 \ll 1$ to dominant order as $1+r+r^2$, $r^2$ and $r^2/2$, respectively. We also find that the reduced density matrix for the cavity field contains the term $(1/\sqrt{2})[(\varepsilon_d^{*}/g)^2 |0\rangle \langle 2| + (\varepsilon_d/g)^2 |2\rangle \langle 0|]$ which produces a contribution proportional to $[(\varepsilon_d^{*}/g)^2 \alpha^2 + (\varepsilon_d/g)^2 (\alpha^{*})^2]e^{-2|\alpha|^2}$ in the Wigner function, breaking once again the symmetry in phase space in a direction determined by the phase of $\varepsilon_d$. In contrast to the two-photon resonance, however, these terms yield a negligible contribution since they are compared against the unity coefficient coming from the vacuum state; higher-order resonances have virtually no effect in the perturbative expansion. Cavity distributions obtained from the numerical solution of Eq.~\eqref{eq:ME1} confirm the prevalence of the vacuum state for the set of ratios $|\varepsilon_d|/g$ used in Fig.~\ref{fig:WssAN}. 
\begin{figure}
\centering
\includegraphics[width=0.485\textwidth]{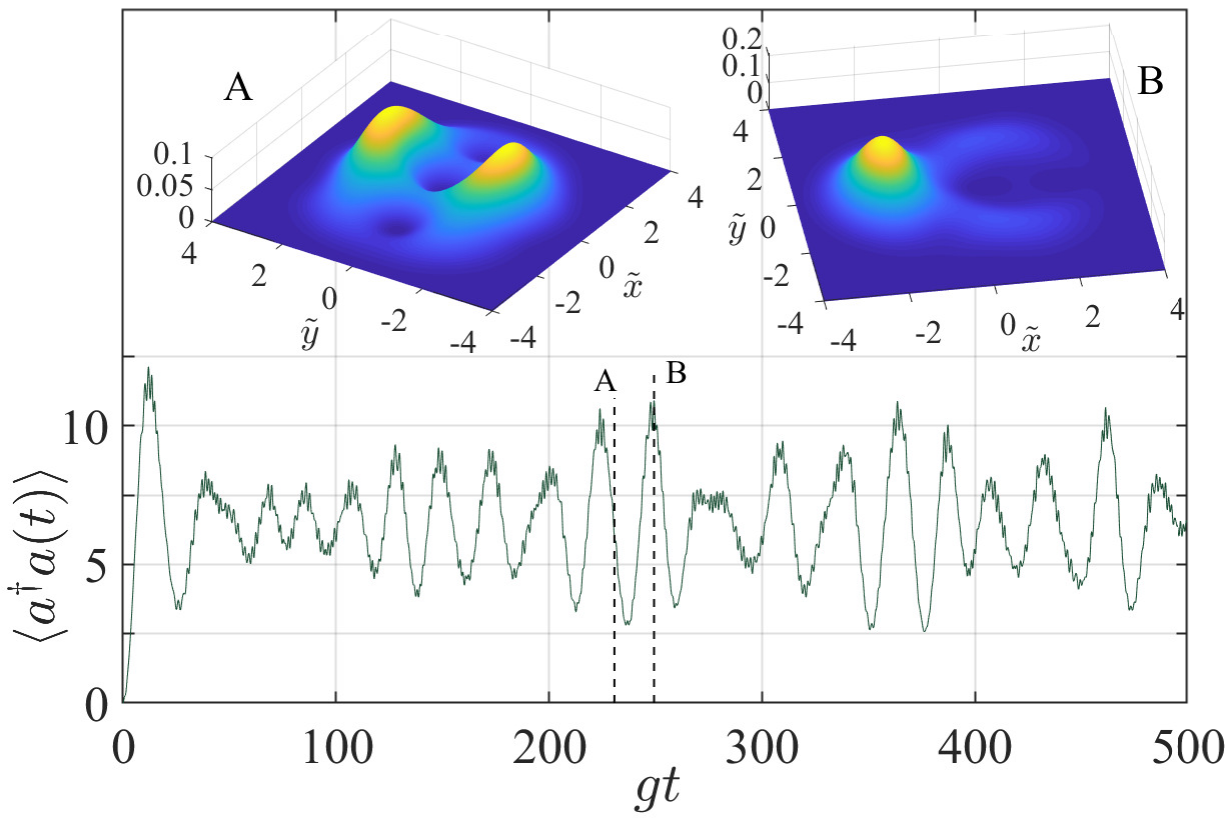}
\caption{{\it Transient bimodality on resonance.} Mean photon number vs. time for $\alpha_0=\sqrt{3}$, calculated from Eqs. (6)--(9) of~\cite{Chough1996} [contrast to Eq. (4.2) of~\cite{Narozhny1981} where the difference frequencies do not dominate]. The two insets depict surface plots of the $Q$ function $Q(\tilde{x} + i \tilde{y})$ of the intracavity field, calculated from Eq.~\eqref{eq:Qz} [with the series truncated at finite values $m=M$ and $n=N$, exceeding $2\alpha_0(\alpha_0+1)$] in the transformed frame for the times $gt=230.67, 249.3$ for A, B, respectively, as marked by the two vertical lines. At the latter time (B), the photon number reaches its maximum value following the initial overshoot [contrast to the dispersion in the distribution at the first revival time in Fig. 3 of~\cite{Chough1996} for $\alpha_0=15$].}
\label{fig:Qres}
\end{figure}

To obtain a non-perturbative analytical result involving transient bimodality at the few-photon level, we make a detour to briefly visit the topic of collapse and revival in the JC model, and consider a closely related {\it non-dissipative} variant where now the two-level atom is coherently driven instead of the cavity mode (for the open version of the problem, the equivalence between atom and cavity driving is elucidated in~\cite{Alsing1991}). The interaction-picture Hamiltonian for this system is~\cite{Chough1996} 
\begin{equation}
H=i g (\sigma_{+}a-\sigma_{-}a^{\dagger}) + i\varepsilon^{\prime}_d (\sigma_{+}-\sigma_{-}),
\end{equation}
where $\varepsilon^{\prime}_d$ is the amplitude of the external field, assumed to be real. Introducing the displacement 
\begin{equation}\label{eq:disptran}
D^{\dagger}(\alpha_0) a D(\alpha_0)=a + \alpha_0, \quad \quad \alpha_0\equiv\varepsilon_d^{\prime}/g,    
\end{equation}
with $D(\alpha_0) \equiv \exp[\alpha_0(a-a^{\dagger})]$, we obtain
\begin{equation}
\tilde{H} \equiv D^{\dagger}(\alpha_0) H D(\alpha_0)= i\hbar g (\sigma_{+}a-\sigma_{-}a^{\dagger}).
\end{equation}
The system state can be written as a sum of two orthogonal contributions, representing excitations of almost orthogonal anharmonic oscillator modes,
\begin{equation}
 \ket{\tilde{\psi}(t)}=(1/\sqrt{2})(\ket{\tilde{U}(t)} + \ket{\tilde{L}(t)})
\end{equation}
with the two orthogonal states expressed as
\begin{equation}\label{eq:Xievol}
 \ket{\tilde{\Xi}(t)}=\sum_{n=1}^{\infty}c_{n-1} e^{\mp i\sqrt{n}t} \ket{\xi_n},
\end{equation}
where $\Xi=(U, L)$. The coefficients $c_{n}\equiv e^{-\alpha_0^2/2}(\alpha_0^n/\sqrt{n!})$ come from the familiar expansion of a coherent state into the Fock-state basis. Here, though [for $n \geq 1$ and adopting a slightly different notation to Eqs.~\eqref{eq:4levels}],
\begin{equation}
 \begin{aligned}
\ket{u_n}&=(1/\sqrt{2})(\ket{n-1}\ket{+}+i\ket{n}\ket{-}), \\
\ket{l_n}&=(1/\sqrt{2})(\ket{n-1}\ket{+}-i\ket{n}\ket{-}), 
 \end{aligned}
\end{equation}
are the excited-state doublets of the JC ladder (with $\ket{0}\otimes\ket{-}$ the ground state).

To visualize the evolution in the phase space we derive the $Q$ function for the intracavity field in the transformed frame, defined as $Q(\tilde{z},\tilde{z}^{*},t)=(1/\pi)[\langle\tilde{z}|\tilde{\psi(t)}\rangle \langle \tilde{\psi(t)}|\tilde{z} \rangle]$ and normalized to unity, from the pure state of Eq.~\eqref{eq:Xievol}. The resulting expression reads ($\tilde{z}=\tilde{x} + i\tilde{y}$)
\begin{widetext}
\begin{align}\label{eq:Qz}
Q(\tilde{z}, \tilde{z}^{*},t)=\frac{1}{2\pi}&\sum_{n=0}^{\infty} \sum_{m=0}^{\infty}\Big(d_n d_m^{*} c_n c_m \big\{\cos[gt(\sqrt{n+1}-\sqrt{m+1})]+ \cos[gt(\sqrt{n+1}+\sqrt{m+1})]\big\}\notag\\
 &+ d_{n+1}d_{m+1}^{*}c_n c_m \big\{\cos[gt(\sqrt{n+1}-\sqrt{m+1})] - \cos[gt(\sqrt{n+1}+\sqrt{m+1})]\big\}\Big),
\end{align}
\end{widetext}
where $d_{n}\equiv e^{-|\tilde{z}|^2/2}(\tilde{z}^n/\sqrt{n!})$. The evolving state of Eq.~\eqref{eq:Xievol} identifies two counter-propagating wavepackets depicted in the two insets of Fig.~\ref{fig:Qres}, which shows the evolution of the intracavity photon number $\braket{a^{\dagger}a(t)}$ (in the original reference frame) from the initial state $\ket{0} \otimes \ket{+}$. The two peaks observed in inset A are located at $\tilde{z}_A=\tilde{x}_A+i\tilde{y}_A \approx 0.28 \pm i 1.70$, with $|\tilde{z}_A|^2$ being very close to $\alpha_0^2=3$. Remnants of these peaks, depicted in the inset B, are still visible at the ``revival'' time when the peak is located at $\tilde{z}_B \approx -\sqrt{3}i$. We note that for high excitation ($\alpha_0^2 \gg 1$), where the time evolution of the system state is described by the asymptotic analysis of~\cite{Banacloche1991}, the $Q$ function---depending on the scaled time $gt$---can be simplified as
\begin{equation}
 Q(\tilde{z}, \tilde{z}^{*},t) \approx \frac{1}{\pi}\sum_{n=0}^{\infty}\sum_{m=0}^{\infty}d_n d_m^{*} c_n c_m \cos[gt(\sqrt{n}-\sqrt{m})],
\end{equation}
showing explicitly the two counter-propagating contributions in interference with each other, through the Euler formula applied to the cosine.

\section{Concluding remarks}

To summarize, in this article we have revealed the distinct character of multi-modality accompanying a multiphoton resonance in a regime where quantum dynamics produces a visible departure from the mean-field predictions. For the special case of the driven two-photon resonance, we have shown that the two peaks of the bimodal distribution are bound by the unit circle for every drive amplitude for which the perturbative treatment remains valid, and are not in correspondence with the two-photon amplitude, i.e., a peak of the distribution at a position vector of magnitude equal to $\sqrt{2}$. If bistability is to be regarded as a feature originating from a semiclassical nonlinearity, then it ties well with the Rabi oscillation developing in the intensity correlation function for the forwards scattered photons as the two-photon transition saturates. At the same time, bimodality is present when we operate well within the strong-coupling regime; as the system-size parameter [$\sim (\gamma/g)^2$] of absorptive bistability -- characteristic of a weak-coupling ``thermodynamic limit'' -- tends to zero for given ratios $\varepsilon_d/g$ and $\gamma/(2\kappa)$, the analytical expression for the Wigner function suggests an increasingly symmetric cavity-field distribution ($p_3 \to 1/4$). Moreover, numerical simulations show a bimodal distribution for the parameters of Fig.~\ref{fig:WssAN}(c) (Panel I) but in the absence of spontaneous emission [$\gamma/(2\kappa)=0$]. While bimodality persists in this so-called ``zero system size'' where the relevance of a system-size parameter [$\sim (g/\kappa)^2$] attributed to a strong-coupling ``thermodynamic limit'' emerges, we encounter an asymmetric distribution where most of the excitation probability gathers in the second quadrant (with the main peak located at $\alpha_{m_1}\approx -0.64 +0.39i$). 

On the experimental front, the phase-space profile of the cavity field in the strong-coupling limit will assist the realization of light sources producing bound states of multiple photons~\cite{Chang2014} in conjunction with autocorrelation functions of order higher than second, to provide a clearer indication of the nonclassical character of the emitted light at weak excitation~\cite{Rundquist2014}. Moreover, the direct measurements of the Wigner function close to the origin of the phase space, in the spirit of~\cite{Lutterbach1997}, can prove a powerful tool for assessing the multiphoton quantum nonlinearity associated with photon blockade. The proposal has been implemented in the experiment of~\cite{Nogues2000} for a single-photon field. The latter is relevant for detecting the alternation between positive and negative values of the Wigner function close to the phase-space origin, correlated with the quantum beat. Finally, the scheme of~\cite{Lutterbach1997} allows one to select the region of phase space to be explored, while being insensitive to a low detection efficiency. Such a direct measurement would readily reveal the distinct features of a multiphoton resonance in the JC oscillator, and uncover a critical behavior which is markedly different than the operation on resonance. 

\begin{acknowledgments}
I acknowledge financial support by the Swedish Research Council (VR) in conjunction with the Knut and Alice Wallenberg foundation (KAW).
\end{acknowledgments}

\bibliography{bibliography}

\end{document}